\documentclass[twocolumn,aps,showpacs,final]{revtex4}

\usepackage{mathptmx,amsmath,amssymb}
\usepackage[polish,english]{babel}
\usepackage[latin1]{inputenc}
\usepackage{graphicx}
\usepackage{times}
\usepackage{bbold}
\usepackage[T1]{fontenc}
\usepackage{epsfig}
\usepackage{epstopdf}
\usepackage{bbold}
\usepackage{amssymb}

\def\lab#1{\label{eq:#1}}

\def\br{\begin{eqnarray}}
\def\er{\end{eqnarray}}
\def\be{\begin{equation}}
\def\ee{\end{equation}}
\def\({\left(}
\def\){\right)}

\def\rlx{\relax\leavevmode}
\def\IR{\rlx\hbox{\rm I\kern-.18em R}}

\def\vt{\vartheta}
\def\u2{\mid u\mid^2}
\newcommand{\sbr}[2]{\left\lbrack\,{#1}\, ,\,{#2}\,\right\rbrack}

\begin{document}

\title{Potentials and the vortex solutions in the $CP^N$ Skyrme-Faddeev model}
\author{Yuki~Amari$^{a}$}
\email{amalyamary@gmail.com}
\author{Pawe\l~Klimas$^{b}$}
\email{pawel.klimas@ufsc.br}
\author{Nobuyuki Sawado$^{a}$}
\email{sawado@ph.noda.tus.ac.jp}
\author{Yuta~Tamaki$^{a}$}
\email{mojyamojya.sax.0313@gmail.com}

\vspace{.5 in}
\small

\affiliation{
$^a$ Department of Physics, Tokyo University of Science,\\
 Noda, Chiba 278-8510, Japan \\
$^{b}$ Universidade Federal de Santa Catarina, Trindade, 88040-900, Florian\'opolis, SC, Brazil
}

\date{\today}

\begin{abstract}
The extended Skyrme-Faddeev model possesses vortex solutions in a (3+1) dimensional Min\-kow\-ski space-time
with target space $CP^N$. They have finite energy per unit of length and contain waves pro\-pa\-ga\-ting along vortices with 
the speed of light. 
We introduce various types of the potentials which correspond with holomorphic solutions of the integrable sector 
and also with several numerical solutions outside of this sector.
The presented solutions constitute a strong indication that the current model contains large class of solutions with 
much wider range of coupling constants than the previously known exact solution. 
\end{abstract}


\maketitle 

\section{Introduction}

The Skyrme-Faddeev model is an example of a field theory that supports the finite-energy knotted solitons. The significance of this model has increased noticeably when it has been conjectured that the model can be seen as a low-energy effective classical model of the underlying Yang-Mills theory \cite{sf}.  Similarly to many other models \cite{coleman} the classical soliton solutions of the Skyrme-Faddeev model can play a role of adequate normal models useful in description of the strong coupling sector of the Yang-Mills theory. The exact soliton (vortex) solution of the model has been found within the integrable sector \cite{vortexlaf}. Such a sector exists in the version of the model that is an extension of the standard Skyrme-Faddeev model obtained by including some quartic term different to the Skyrme term. The study of the extended models have been originally motivated by the results of the analysis of the Wilsonian action of the $SU(2)$ Yang-Mills theory \cite{gies}. It has been shown that also in the case of the complex projective target space $CP^N$ the extended Skyrme-Faddeev model 
 (in which it has been imposed some special constraints for the parameters of the model) 
possesses an exact soliton solutions in the integrable sector \cite{fk,fkz}. It has not been clear until now if the presence of the solutions in the considered model is related to the particular choice of the coupling constants like in the case of the exact solution or it is rather a general property of the model. In order to answer this question one needs to construct some other solutions than the exact ones.
The aim of this paper is to investigate the existence of the solutions of the model inside/outside the integrable sector, 
especially in absence of some particular relations between coupling constants. We show that such solutions exist. 
The key role is playing by the potential which usually work as a stabilizer for the solution. 
In the present paper the potential appears in the context of exact holomorphic solutions, as 
it was presented in the case of $CP^1$ related model~\cite{fjst,Sawado:2013yza}, and also for solutions from a non-holomorphic sector.   
We conclude that the exact solution appears as a particular solution belonging to the wider class of solutions of the model.

The study of such models is promising and could be important for understanding some aspects of the strong coupling sector of the Yang-Mills theory.

\section{The formulation of the model}

The Skyrme-Faddeev model and its extensions on the $CP^1$ target space are usually expressed in terms of the real unit vector $\vec n$. The dimension of a target space is simply related to the number of degrees of freedom of the model. For instance, the model with the $CP^1$ target space has only two independent degrees of freedom. In order to add more degrees of freedom one can consider some higher dimensional target spaces.  The target space (coset space) in the case of some higher dimensional $SU(N)$ Lie groups, i.e. $N>2$, can be chosen in several nonequivalent ways. 

Recently it has been proposed some formulation of the extended Skyrme-Faddeev model on the $CP^N$ target space \cite{fk}. The coset space $CP^N=SU(N+1)/SU(N)\otimes U(1)$ is an example of a symmetric space and it can be naturally parameterized in terms of so called {\it principal variable} $X(g)=g\sigma(g)^{-1}$, with $g\in SU(N+1)$ and $\sigma$ being the order two automorphism under which the subgroup $SU(N)\otimes U(1)$ is invariant i.e. $\sigma(h)=h$ for $h\in SU(N)\otimes U(1)$. The principal coordinate $X(g)$ defined above satisfies $X(gh)=X(g)$.
 
We shall consider the field theory in $(3+1)$ dimensions defined by the Lagrangian
\begin{eqnarray}
&&{\cal L}= -\frac{M^2}{2}{\rm Tr}\(X^{-1}\partial_{\mu}X\)^2
+\frac{1}{e^2}{\rm
  Tr}\(\sbr{X^{-1}\partial_{\mu}X}{X^{-1}\partial_{\nu}X}\)^2 \nonumber\\
&&+
\frac{\beta}{2}\left[{\rm Tr}\(X^{-1}\partial_{\mu}X\)^2\right]^2
+\gamma\left[{\rm Tr}\(X^{-1}\partial_{\mu}X\;\;
X^{-1}\partial_{\nu}X\)\right]^2
\nonumber \\
&&-\mu^2V
\label{actionx}
\end{eqnarray}
where $M$ is a coupling constant with dimension of mass whereas the coupling constants $e^{-2}$, $\beta$, $\gamma$ are dimensionless. 
The first term is quadratic in $X$ and corresponds with the Lagrangian of the $CP^N$ model. The quartic term proportional to $e^{-2}$ 
is the Skyrme term whereas other quartic terms constitute the extension of the standard Skyrme-Faddeev model. 
The novelty of the model (\ref{actionx}) comparing with that introduced in \cite{fk} is the presence of the potential $V$. 
In recent studies, several potentials have been introduced for the planar Skyrme-type model
~\cite{wer,Jaykka:2010bq,Nitta}. 
It was shown that the extended $CP^1$ Skyrme-Faddeev model in (3+1) dimensions possesses some non-holomorphic 
solutions that do not belong to the integrable sector \cite{fjst}. Since the extended Skyrme-Faddeev model 
on the $CP^N$ target space also possesses the integrable sector as well as the exact vortex solutions the 
natural question is if there exist any solutions that do not belong to the integrable sector? In similarity 
to the paper \cite{fjst} we study such a possibility in the presence of the potential. As it has been 
explained below such solutions can be obtained numerically for some choice of the potential. 

\subsection{The parametrization}
Let us shortly discuss the parameterization of the model. According to the previous paper \cite{fk} 
one can parametrize the model in terms of $N$ complex fields $u_i$, where $i=1,\ldots,N$. 
Assuming $(N+1)$-dimensional defining representation where the $SU(N+1)$ valued element $g$ is of the form
\br
g\equiv \frac{1}{\vt}\,\(\begin{array}{cc}
\Delta&i\,u\\
i\,u^{\dagger}&1
\end{array}\) \qquad\quad \vt\equiv\sqrt{1+u^{\dagger}\cdot u}
\lab{gdef}
\er
and where $\Delta$ is the hermitian $N\times N$-matrix
\br
\Delta_{ij}=\vt\,\delta_{ij}-\frac{u_i\,u_j^*}{1+\vt}
~~\mbox{\rm which satisfies}
~~\Delta\cdot u= u~~\mbox{\rm and}~~u^{\dagger}\cdot \Delta= u^{\dagger}.
\label{deltadef}
\nonumber 
\er
The principal variable $X(g)=g\sigma(g)^{-1}$ takes the form
\begin{eqnarray}
X(g)=g^2=
\(\begin{array}{cc}
I_{N\times N} & 0 \nonumber \\
0 & -1 \nonumber 
\end{array}\)
+
\frac{2}{\vt^2}\(\begin{array}{cc}
-u\otimes u^\dagger & iu \nonumber \\
iu^\dagger & 1 \nonumber 
\end{array}\)
\end{eqnarray}
and the Lagrangian (\ref{actionx}) reads 
\be
{\cal L}=
-\frac{1}{2} \Bigl[M^2 \eta_{\mu\nu}+C_{\mu\nu}\Bigr]\tau^{\nu\mu}-\mu^2V
\label{actioncmunu}
\ee
where the symbols $C_{\mu\nu}$ and $\tau_{\mu\nu}$ are defined as follows 
\begin{eqnarray}
C_{\mu\nu}:= M^2\eta_{\mu\nu}-\frac{4}{e^2}\Bigl[(\beta e^2-1)\tau_{\rho}^{\rho}\eta_{\mu\nu}
\nonumber \\
+(\gamma e^2-1)\tau_{\mu\nu}+(\gamma e^2+2)\tau_{\nu\mu}\Bigr],
\label{cmunudef}
\end{eqnarray}
\be
\tau_{\mu\nu}:=-\frac{4}{\vartheta^4}\left[\vartheta^2
\partial_{\nu}u^{\dagger}\cdot \partial_{\mu}u-(\partial_{\nu}u^{\dagger}\cdot u)(u^{\dagger}\cdot\partial_{\mu}u) \right].
\ee
A variation with respect to $u_i^*$ leads to the equations which can be cast in the form
\begin{eqnarray}
&&(1+u^{\dagger}\cdot u)\partial^{\mu}(C_{\mu\nu}\partial^{\nu}u_i) \nonumber \\
&&-C_{\mu\nu}\left[(u^{\dagger}\cdot\partial^{\mu}u)\partial^{\nu}u_i+(u^{\dagger}\cdot\partial^{\nu}u)\partial^{\mu}u_i\right]\nonumber\\
&&+\frac{\mu^2}{4}(1+u^{\dagger}\cdot u)^2\sum_{k=1}^{N}\left[(\delta_{ik}+u_iu^*_k)\frac{\delta V}{\delta u^*_k}\right]=0.
\label{eom1}
\end{eqnarray}
where we have  already multiplied the resultining equations by  inverse of $\Delta^2_{ij}$ i.e. $\Delta^{-2}_{ij}=\frac{1}{1+u^{\dagger}\cdot u}(\delta_{ij}+u_i u^{*}_j)$. 
We shall discuss some examples of the potential in the further part of the paper. In the simplest case when the potential is a function of absolute values of the fields $V(|u_1|^2,\ldots,|u_N|^2)$ the contribution from the potential becomes
$$
\sum_{k=1}^{N}\left[(\delta_{ik}+u_iu^*_k)\frac{\delta V}{\delta u^*_k}\right]=u_i\left[\frac{\delta V}{\delta |u_i|^2}+\sum_{k=1}^N|u_k|^2\frac{\delta V}{\delta |u_k|^2}\right].
$$
We introduce the dimensionless coordinates $(t,\rho,\varphi,z)$ defined as
\br
x^0=r_0t,\quad x^1=r_0\rho\cos\varphi,\quad x^2=r_0\rho\sin\varphi\quad x^3=r_0 z
\er
where the length scale $r_0$ is defined in terms of coupling constants $M^2$ and $e^2$ i.e.
$$
r_0^2=-\frac{4}{M^2e^2}
$$
and the light speed is $c=1$ in the natural units. The linear element $ds^2$ reads
$$
ds^2=r_0^2(dt^2-dz^2-d\rho^2-\rho^2d\varphi^2).
$$
The family of exact vortex solutions has been found for the model without potential 
$\mu^2=0$ where in addition the coupling constants satisfy the condition $\beta e^2+\gamma e^2=2$. 
The exact solutions have the form of vortices which depend on some specific combination of the coordinates 
i.e. one light-cone coordinate $x^3+x^0$ and one complex coordinate $x^1+ix^2$. The functions $u_i(x^3+x^0,x^1+ix^2)$ 
satisfy the zero curvature condition $\partial_{\mu}u_i\partial^{\mu}u_j=0$ for all $i,j=1,\ldots,N$ and
 therefore one can construct the infinite set of conserved currents. 

We shall consider the following ansatz
\br
u_j=f_j(\rho)e^{i(n_j\varphi+k_j\psi(w))}\label{ansatz}
\er
where $\psi(w)$ is a real function of the light-cone coordinate and $f_i(\rho)$ are real-valued functions. The constants $n_i$ form the set of integer numbers and $k_i$ are some real constants. 
The holomorphic solutions, which belong to the integrable sector, is of the form
\begin{eqnarray}
f_i(\rho)=c_i\rho^{n_i}
\label{holomorphic}
\end{eqnarray}
where $c_i$ are some real (in general complex) free constants.  We define two diagonal matrices
\br
\lambda\equiv{\rm diag}(n_1,\ldots,n_N),\qquad\sigma\equiv{\rm diag}(k_1,\ldots,k_N).
\er 
in order to simplify the form of some formulas below. In matrix notation the ansatz reads $u=f(\rho)\exp{[i(\lambda\varphi+\sigma\psi(w))]}$ where $w$ is either $z+t$ or $z-t$. 
The expressions $\tau_{\mu\nu}$ have the following form
\br
&&\tau_{\rho\rho}\equiv\theta(\rho)=-\frac{4}{\vt^4}\,\left[\vt^2\,f'^T.f'-(f'^T.f)(f^T.f')\right]\nonumber\\
&&\tau_{\varphi\varphi}\equiv\omega(\rho)=-\frac{4}{\vt^4}\,\left[\vt^2\,f^T.\lambda^2.f-(f^T.\lambda.f)^2\right]\nonumber\\
&&\tau_{\varphi\rho}=-\tau_{\rho\varphi}\equiv i\zeta(\rho) \nonumber \\
&&~~~~~~~~\zeta(\rho)=-\frac{4}{\vt^4}\,\left[\vt^2\,f'^T.\lambda.f-(f^T.\lambda.f)(f'^T.f)\right] \nonumber\\
&&\tau_{t\rho}=-\tau_{\rho t}\equiv (i\partial_t\psi)\xi(\rho),~~~~\tau_{z\rho}=-\tau_{\rho z}\equiv (i\partial_z\psi)\xi(\rho)\nonumber \\
&&~~~~~~~~\xi(\rho)=-\frac{4}{\vt^4}\,\left[\vt^2\,f'^T.\sigma.f-(f^T.\sigma.f)(f'^T.f)\right] \nonumber\\
&&\tau_{t\varphi}=\tau_{\varphi t}\equiv (\partial_t\psi)\eta(\rho),~~~~\tau_{z\varphi}=\tau_{\varphi z}\equiv (\partial_z\psi)\eta(\rho)\nonumber \\
&&~~~~~~~~\eta(\rho)=-\frac{4}{\vt^4}\,\left[\vt^2\,f^T.\lambda.\sigma.f-(f^T.\lambda.f)(f^T.\sigma.f)\right]\nonumber\\
&&\tau_{tt}\equiv (\partial_t\psi)^2\chi(\rho),~~~~\tau_{zz}\equiv (\partial_z\psi)^2\chi(\rho) \nonumber \\
&&\tau_{tz}=\tau_{zt}\equiv (\partial_t\psi)(\partial_z\psi)\chi(\rho) \nonumber \\
&&~~~~~~~~\chi(\rho)=-\frac{4}{\vt^4}\,\left[\vt^2\,f^T.\sigma^2.f-(f^T.\sigma.f)(f^T.\sigma.f)\right] \nonumber
\er
where derivative with respect to $\rho$ is denoted by $\frac{d}{d\rho}='$ and $T$ stands for matrix transposition. The equations of motion written in dimensionless coordinates take the form
\br
&&(1+f^T.f)\left[\frac{1}{\rho}\left(\rho\,\tilde{C}_{\rho\rho}f'_k\right)'+\frac{i}{\rho}\left(\frac{\tilde{C}_{\rho\varphi}}{\rho}\right)'(\lambda.f)_k-\frac{1}{\rho^4}\tilde{C}_{\varphi\varphi}(\lambda^2.f)_k\right]\nonumber\\
&&-2\left[\tilde{C}_{\rho\rho}(f^T.f')f'_k-\frac{1}{\rho^4}\tilde{C}_{\varphi\varphi}(f^T.\lambda.f)(\lambda.f)_k\right]\nonumber\\
&&+\tilde{\mu}^2\frac{f_k}{4}(1+f^{T}.f)^2\left[\frac{\delta V}{\delta f_k^2}+\sum_{i=1}^Nf_i^2\frac{\delta V}{\delta f_i^2}\right]=0
\label{equationr}
\er
for each $k=1,\ldots,N$, where we have introduced the symbols 
$\tilde {C}_{\mu\nu}:= \frac{1}{M^2}C_{\mu\nu}$, and also $\tilde{\mu}^2:=\frac{r_0^2}{M^2}\mu^2$. 
The components $\tilde{C}_{\mu\nu}$ which appear in the equations of motion read
\br
&&\tilde{C}_{\rho\rho}=-1+(\beta e^2-1)\left(\theta+\frac{\omega}{\rho^2}\right)+(2\gamma e^2+1)\theta\nonumber\\
&&\tilde{C}_{\varphi\varphi}=-\rho^2+\rho^2(\beta e^2-1)\left(\theta+\frac{\omega}{\rho^2}\right)+(2\gamma e^2+1)\omega\nonumber\\
&&\tilde{C}_{\varphi\rho}=-\tilde{C}_{\rho\varphi}=-3i\zeta\,.
\er

%

\subsection{The energy}
The Hamiltonian density being a Legendre transform of the Lagrangian density (\ref{actioncmunu}) is defined as follows
\br
\mathcal{H}&:=&\frac{\delta\, {\cal L}}{\delta\,\partial_{0}u_i}\,\partial_0 u_i 
+\frac{\delta\, {\cal L}}{\delta\,\partial_{0}u_i^*}\,\partial_0 u_i^* 
- {\cal L}.\label{defhamiltonian}
\er
The resulting Lagrangian and Hamiltonian densities taken for the solution (\ref{ansatz}) depend only on the coordinates $\rho$ and either $z+t$ or $z-t$. For the Lagrangian density one gets
\br
\mathcal{L}&=&\frac{M^2}{r_0^2}\left(\theta+\frac{\omega}{\rho^2}\right)+\frac{2}{r_0^4e^2}(\beta e^2+\gamma e^2-2)\left(\theta+\frac{\omega}{\rho^2}\right)^2
\nonumber \\
&+&\frac{4}{r_0^4e^2}(\gamma e^2-1)\frac{\zeta^2-\theta\omega}{\rho^2} \nonumber \\
&+&\frac{2}{r_0^4e^2}(\gamma e^2+2)\left(\theta^2+\frac{\omega^2}{\rho^4}-2\frac{\zeta^2}{\rho^2}\right)
-\mu^2V\,
\er
where a term proportional to $M^2$ is just the $CP^N$ Lagrangian and terms  proportional to $\gamma e^2-1$ and $\gamma e^2+2$, vanish for the holomorphic solutions (\ref{holomorphic}) 
since the constraint $\partial_{\mu}u_i\partial^{\mu}u_j=0$ leads to the relation 
$f'_j(\rho)=\frac{n_j}{\rho} f_j(\rho)$ resulting in equalities $\theta=\frac{\omega}{\rho^2}=\frac{\zeta}{\rho}$ 
and $\xi=\frac{\eta}{\rho}$. 

The Hamiltonian density (\ref{defhamiltonian}) considered for (\ref{ansatz}) contains several terms. We present it in dimensionless form (in the unit of $-M^4e^2/4$)
\br
\mathcal{H}=\sum_{j=1}^7\mathcal{H}_j
\label{hamiltonian}
\er
where the  components $\mathcal{H}_j$  are given by
\begin{eqnarray}
&&\mathcal{H}_1=-\left(\theta+\frac{\omega}{\rho^2}\right)\,,~~~~
\mathcal{H}_2=-2\left(\frac{d\psi}{dw}\right)^2\chi\,,\nonumber \\
&&\mathcal{H}_3=2\left(\frac{d\psi}{dw}\right)^2\left\{(\beta e^2-1)\left(\theta+\frac{\omega}{\rho^2}\right)\chi
+(\gamma e^2-1)\frac{2\eta^2}{\rho^2}\right\}
\nonumber \\
&&\mathcal{H}_4=\frac{1}{2}(\beta e^2+\gamma e^2-2)\left(\theta+\frac{\omega}{\rho^2}\right)^2
\nonumber \\
&&\mathcal{H}_5=(\gamma e^2-1)\frac{\zeta^2-\theta\omega}{\rho^2}
+\frac{1}{2}(\gamma e^2+2)\left(\theta^2+\frac{\omega^2}{\rho^4}-2\frac{\zeta^2}{\rho^2}\right)
\nonumber \\
&&\mathcal{H}_6=-6\left(\frac{d\psi}{dw}\right)^2\left(\xi^2-\frac{\eta^2}{\rho^2}\right)\,,~~~~
\mathcal{H}_7=\tilde{\mu}^2 V.
\label{hamiltonian_def}
\end{eqnarray}
We have split the Hamiltonian density in order to make explicit the terms that were present 
in the earlier study of the holomorphic vortex solutions i.e $\mathcal{H}_1$, 
$\mathcal{H}_2$ and $\mathcal{H}_3$. The term $\mathcal{H}_4$ and also the potential 
term $\mathcal{H}_7$ were absent in previous considerations. 
They did not appear due to the constraint imposed on the coupling constants $\beta e^2+\gamma e^2=2$
and also due to the integrability condition. 
The terms $\mathcal{H}_5,\mathcal{H}_6$ are zero for the holomorphic solutions. 
Note that for static ($w$-independent) vortex solutions, $\mathcal{H}_2,\mathcal{H}_3,\mathcal{H}_6$ reduce to zero
and then, only the terms $\mathcal{H}_4,\mathcal{H}_5,\mathcal{H}_7$ could be meaningful for
the Derrick's scaling argument. 

\subsection{The topological property}
According to the discussions in \cite{D'Adda:1978uc} and also in \cite{fk}, we can define the 
topological charge in the present model. The field $u_i$ provide a mapping from  $x^1x^2$ plane 
into $CP^N$. However, for the finiteness of the energy, the field goes to a constant. Then the 
plane should be compactified into $S^2$ and the solutions define the mapping $S^2\to CP^N$
which is classified into the homotopy classies of $\pi_2(CP^N)$.  There exists a theorem 
describing in \cite{D'Adda:1978uc}, $\pi_2(G/H)=\pi_1(H)_G$ where $\pi_1(H)_G$ is the subset 
of $\pi_1(H)$ formed by closed paths in $H$ which can be contracted to a point in $G$. Thus, 
in the present case, the topological charges are given by 
\begin{eqnarray}
\pi_1(SU(N)\otimes U(1))_{SU(N+1)}.
\end{eqnarray}
As discussed in \cite{D'Adda:1978uc},\cite{fk}, the topological charges are equal to the 
number of poles of $u_i$, including those at infinity. And then, it can be obtained as
\begin{eqnarray}
Q_{\rm top}=n_{\rm max}+|n_{\rm min}|
\end{eqnarray}
where the highest positive integer in the set $n_i, i=1,2,\cdots,N$ and $n_{\rm min}$ is the 
lowest negative integer in the same set. 

\section{Reduction to the integrable $CP^1$ sector}

In \cite{fjst}, one of us have claimed that for some special choice of the potential $V$ there exist analytical solutions of the $CP^1$ Skyrme-Faddeev type model for all topological charges. It turns out that for a $CP^N$ version of the extended Skyrme-Faddeev model there exists sectors such that the model reduces to the $CP^1$ version. One can expect that when reduction occurs the model possesses holomorphic solutions for the appropriate choice of the potential. In current section we shall study this problem in details.

We are interested in a case such that the quartic term proportional to $\beta e^2+\gamma e^2-2$ does contribute to equations of motion and solutions of those equations are of the form
\begin{eqnarray}
u_k=c_k \rho^{n_k}e^{i(n_k\varphi+k\psi(w))}
\label{solution_hol}
\end{eqnarray}
with $c_k$ being some real constant parameters. The form of solution solves the  zero-curvature condition $\partial_{\mu}u_i\partial^{\mu}u_j=0$. The problem we have to face is in fact an inverse problem i.e. we shall derive the form of the potential for a given solution.
Let us observe that the equations of motion for the solution (\ref{solution_hol}) simplifies a lot taking the form
\br
\tilde\nu^2\frac{n_i}{\rho^4}\frac{R(\rho)}{(1+u^{\dagger}\cdot u)^4}=\tilde \mu^2\left[\frac{\delta V}{\delta |u_i|^2}+\sum_{k=1}^N|u_k|^2\frac{\delta V}{\delta |u_k|^2}\right]\label{eom}
\er
where we have denoted $\tilde\nu^2:=64(\beta e^2+\gamma e^2-2)$ and 
$$
R(\rho):=\sum_{i=1}^NA_i|u_i|^2+\sum_{i,j=1}^NB_{ij}|u_i|^2|u_j|^2+\sum_{i,j,k=1}^NC_{ijk}|u_i|^2|u_j|^2|u_k|^2
$$
where the coefficients are functions of integers $n_i$
\br
&&A_i=n_i^2(n_i-1)\nonumber\\
&&B_{ij}=n_i^2(n_i-2n_j-1)+\frac{1}{2}(n_i-n_j)^2(n_i+n_j-1)\nonumber\\
&&C_{ijk}=\frac{1}{2}(n_i-n_j)^2(n_i+n_j-2n_k-1).\nonumber
\er
One can get rid of the denominator $(1+u^{\dagger}\cdot u)^{4}$ on both sides of the field equations substituting 
$$
V=\frac{W}{(1+u^{\dagger}\cdot u)^4}.
$$ 
The resulting set of equations $i=1,\ldots,N$ has the following form
\br
\frac{\tilde\nu^2}{\tilde\mu^2}\frac{n_i}{\rho^4}R(\rho)=\frac{\delta W}{\delta |u_i|^2}+\sum_{k=1}^N|u_k|^2\frac{\delta W}{\delta |u_k|^2}-4W.\label{eq1}
\er
The function $W$ must have such a form that the complete system of $N$ equations (\ref{eq1}) holds. Such a function $W$, satisfying the set of equation with arbitrary integers $(n_1,n_2,\ldots,n_N)$, would be a true generalization of the problem of exact solutions to the $CP^N$ case. 

\subsection{The potential}
Instead of solving this (still open) question we shall study the case of reduction with only one non-zero integer $n$. We assume that the first $K$ integers $n_i$ are equal $n_1=n_2=\ldots n_K\equiv n$ and rest of them vanish $n_{K+1}=\ldots =n_N\equiv 0$.
The set of reduced equations (\ref{eq1}) takes the form
\br
\left\{
\begin{array}{rcl}
\frac{\tilde \nu^2}{\tilde\mu^2}\frac{n}{\rho^4}R(\rho)-\left(\frac{\delta W}{\delta |u_i|^2}+\sum_{k=1}^N|u_k|^2\frac{\delta W}{\delta |u_k|^2}-4W\right)&=&0 \\
\frac{\delta W}{\delta |u_i|^2}+\sum_{k=1}^N|u_k|^2\frac{\delta W}{\delta |u_k|^2}-4W&=&0
\end{array}\right.\label{set1}
\er
where the first subset of (\ref{set1}) is labeled by $i=1,\ldots,K$ and the second subset by $i=K+1,\ldots,N$.
The solutions $u_i$ for $c_1=\ldots=c_K\equiv c$ satisfy 
\br
\begin{array}{lll}
&|u_i|^2=:|u|^2=c^2\rho^{2n} &\mbox{for }\quad i=1,\ldots K\\
&|u_i|^2=c_i^2 & \mbox{for }\quad i=K+1,\ldots N
\end{array}\label{restriction}
\er
which implies that the function $R(\rho)$ in (\ref{set1}) simplifies to the form
$$
R(\rho)=\frac{n^3}{2}K^2\left(1+\sum_{k=K+1}^Nc_k^2\right)^2\left[\alpha|u|^2-\delta|u|^4\right]
$$
where the coefficients $\alpha$, $\delta$ read
\br
\alpha:=\frac{2}{K}\left(1-\frac{1}{n}\right)\qquad \delta:=\frac{2}{1+\sum_{k=K+1}^Nc_k^2}\left(1+\frac{1}{n}\right).\label{alphadelta}
\er
The problem can be solved by the method of separation of variables. For this reason we consider the function $W$ in the form of a product of two functions $P$ and $Q$
$$
W=P(|u_1|^2,\ldots,|u_K|^2)Q(|u_{K+1}|^2,\ldots,|u_N|^2)
$$
where the function $P$ is a product
$$
P(|u_1|^2,\ldots,|u_K|^2)=\prod_{l=1}^K|u_l|^{2\tilde\alpha}
$$ 
where the parameter $\tilde\alpha$ is a free constant. Considering that $\frac{\delta P}{\delta |u_j|^2}=\frac{\tilde\alpha}{|u_j|^2}P$ and $\sum_{j=1}^K|u_j|^2\frac{\delta P}{\delta |u_j|^2}=K\tilde\alpha P$ one can show that the second subset $i=K+1,\ldots,N$ of equations (\ref{set1}) reduces to the following one
\br
\frac{\delta Q}{\delta |u_i|^2}+\sum_{j=K+1}^N|u_j|^2\frac{\delta Q}{\delta |u_j|^2}+(K \tilde\alpha-4)Q=0.\label{eq2}
\er  
The equations (\ref{eq2}) considered for $|u_i|^2=c_i^2$ hold for any $i=K+1,\ldots,N$ if the function $Q$ is such that 
\br
\left[\frac{1}{Q}\frac{\delta Q}{\delta |u_i|^2}\right]_{|u_k|^2=c_k^2}=\tilde\delta\label{condQ}
\er
where $\tilde\delta$ is another free constant. Consequently the equations (\ref{eq2}) reduce to the relation between constants $\tilde \alpha$ and $\tilde\delta$ 
\br
\tilde\delta=\frac{4-K\tilde\alpha}{1+\sum_{j=K+1}^Nc_j^2}\label{delta}.
\er 
For the first subset $i=1,\ldots,K$ of the equations (\ref{set1}) one gets 
$$
\frac{\tilde \nu^2}{\tilde\mu^2}\frac{n}{\rho^4}R(\rho)=\left[\frac{\tilde\alpha}{|u_i|^2}+K\tilde\alpha+\sum_{j=K+1}^N|u_j|^2\frac{1}{Q}
\frac{\delta Q}{\delta |u_j|^2}-4\right]PQ
$$
where the rhs of this formula is taken for $|u_j|^2$ given by (\ref{restriction}). It follws that the last formula in fact became
$$
\frac{\tilde \nu^2}{\tilde\mu^2}\frac{n}{\rho^4}R(\rho)=\left[\frac{\tilde\alpha}{|u|^2}-\tilde\delta\right]PQ.
$$
where $P=|u|^{2K\tilde\alpha}$, $Q=Q(c_{K+1}^2,\ldots,c_N^2)$. In the last step we have made use of the relation $(\ref{delta})$. Putting all results together we obtain 
\br
&&\frac{\tilde \nu^2}{\tilde\mu^2}\frac{n}{\rho^4}\frac{n^3}{2}K^2\left(1+\sum_{k=K+1}^Nc_k^2\right)^2\left[\alpha|u|^2-\delta|u|^4\right]\nonumber\\
&&-\left[\tilde\alpha (|u|^2)^{K\tilde\alpha-1}-\tilde\delta (|u|^2)^{K\tilde\alpha}\right]Q(c_{K+1}^2,\ldots,c_N^2)=0.\nonumber
\er
One can conclude from the last equation that both free constants $\tilde\alpha$ and $\tilde\delta$ have to be fixed by $\tilde\alpha=\alpha$ and $\tilde\delta=\delta$. In such a case one gets 
$$
\left[\alpha (|u|^2)^{K\alpha-1}-\delta (|u|^2)^{K\alpha}\right]=\frac{1}{c^{\frac{4}{n}}\rho^4}\left[\alpha |u|^2-\delta |u|^4\right]
$$
and all equations take the form of relation between constants which fixes value of $\tilde\mu^2$ in terms of other constants
\br
\tilde\mu^2=32(\beta e^2+\gamma e^2-2)c^{\frac{4}{n}}n^4K^2\frac{\left(1+\sum_{j=K+1}^Nc_j^2\right)^2}{Q(c_{K+1}^2,\ldots,c_N^2)}.\label{mu}
\er
The only condition that the function $Q$ has to satisfy is that given by (\ref{condQ}) with $\tilde\delta=\delta$. An example of the function $Q$ is given in next subsection.

We conclude from this section that for the potential
$$
V=\frac{(\prod_{l=1}^K|u_l|^2)^{\alpha}Q(|u_{K+1}|,\ldots,|u_N|^2)}{(1+\sum_{j=1}^N|u_j|^2)^4}
$$
where $Q$ satisfy the condition (\ref{condQ}) in the case of reduction to $CP^1$ the model possesses holomorphic solutions in the sector $\beta e^2+\gamma e^2\neq 2$. 

\subsubsection*{An example of the exact potential}
Let us consider the function $Q$ in the form
$$
Q=\left[\prod_{l=K+1}^N(|u_l|^2)^{-\frac{1}{n}}S(|u_{K+1}|^2,\ldots,|u_N|^2)\right]^2.
$$
In such a case the condition (\ref{condQ}) i.e. $
\left[\frac{\delta}{\delta|u_j|^2}\ln Q\right]_{c_k^2}=\delta
$ became
$$
\left[\frac{\delta}{\delta|u_j|^2}\ln S\right]_{c_k^2}=\frac{1}{nc_j^2}+\frac{\delta}{2}.
$$
We shall assume the function $S$ in the form of linear combination of $|u_l|^2$ i.e. $S={\cal A}-\sum_{l=K+1}^N{\cal B}_l|u_l|^2$ which reduce the last condition to the set of equations
\br
&&\left({\cal A}-\sum_{l=K+1}^N{\cal B}_lc_l^2\right)\left(1+\sum_{l=K+1}^Nc_l^2+(n+1)c_j^2\right)\nonumber\\&&+n\left(1+\sum_{l=K+1}^Nc_l^2\right)c_j^2{\cal B}_j=0\nonumber
\er
that have to hold for any $j=K+1,\ldots,N$. All coefficients proportional to $c_j^2$ as well as free terms must vanish independently. Finally it leads to two equations what suggest parametrization containing only two variables $a$ and $b$. Taking $c_j^2{\cal B}_j=a\,c_j^2+b$ we reduce the set of equations to the following one
\br
\left\{
\begin{array}{ll}
&\left(\sum_{l=K+1}^Nc_l^2-n\right)a+(n+1)(N-K)b=(n+1){\cal A}\\
&\left(\sum_{l=K+1}^Nc_l^2\right)a+(N-K-n)b={\cal A}
\end{array}\right.\nonumber
\er
that have solutions in the form
\br
\left\{
\begin{array}{rl}
a&=\frac{(n+1){\cal A}}{-(n+1)+(N-K+1)(1+\sum_{l=K+1}^Nc_l^2)}\\
b&=\frac{(1+\sum_{l=K+1}^Nc_l^2){\cal A}}{-(n+1)+(N-K+1)(1+\sum_{l=K+1}^Nc_l^2)}.
\end{array}\right.\nonumber
\er
For a particular choice of ${\cal A}$ i.e.
\br
{\cal A}=-(n+1)+(N-K+1)(1+\sum_{l=K+1}^Nc_l^2)\label{A}
\er
solutions reduce to $a=n+1$ and $b=1+\sum_{l=K+1}^Nc_l^2$. It follows that
\br
{\cal B}_j=n+1+\frac{1}{c_j^2}\left(1+\sum_{l=K+1}^Nc_l^2\right).\label{B}
\er
The set of coefficients $c_{K+1},\ldots, c_N$ determines the constants ${\cal A}$ and ${\cal B}_j$. From the physical point of view the inverse problem is more interesting i.e. when the potential parameters are free constants. In such a case one has to invert the relations between $c_j$ and ${\cal B}_j$. The parameter ${\cal A}$ is not independent constant since it is determined by values of constants $c_j$. The function $Q$ gets the form 
\br
Q=\left[\prod_{l=K+1}^N(|u_l|^2)^{-\frac{1}{n}}\left({\cal A}-\sum_{l=K+1}^N{\cal B}_l|u_l|^2\right)\right]^2
\er
which implies the formula
$$Q(c_{K+1}^2,\ldots,c_N^2)=n^2(1+\sum_{j=K+1}^Nc_j^2)^2\prod_{l={K+1}}^Nc_l^{-\frac{4}{n}}.$$
Plugging this result to (\ref{mu}) one gets the relation
$$
\tilde\mu^2=32(\beta e^2+\gamma e^2-2)n^2K^2c^{\frac{4}{n}}\prod_{l={K+1}}^Nc_l^{\frac{4}{n}}.
$$

\subsection{The energy of an exact vortex configuration}
In this subsection we present exact expressions for the energy of the vortex configurations being analytical solutions in the model containing potential. To make formulas less complex we shall consider a simplified case such that there is only one free constant $c_k\equiv c$ for all $k=1,\ldots,K$ and $c_k\equiv1$ for $k=K+1,\ldots,N$.  It follows that 
\br
\alpha:=\frac{2}{K}\left(1-\frac{1}{n}\right)\qquad \delta:=\frac{2}{N-K+1}\left(1+\frac{1}{n}\right).\label{param}
\er
The energy per unit of length of the vortex is a sum  $\sum_{k=1}^7E_k$ where  contributions $E_k$ are defined as 
$$
E_k=2\pi M^2\int_0^{\infty}d\rho\rho\mathcal{H}_k.
$$
The first term $E_1$ is purely topological and therefore it is proportional to $|n|$
$$
E_1=8\pi M^2|n|.
$$
The energy $E_2$ can be cast in the form of the sum
$$
E_2=8\pi M^2\psi'^2\frac{1}{|n|\,c^{\frac{2}{n}}}\left[\frac{N-K+1}{K}\right]^{\frac{1}{n}}\sum_{j=1}^3a_jI_j(n)
$$
where the coefficients $a_j$ depend on parameters $k_i$
\br
a_1&:=&\frac{1}{K}\sum_{i=1}^Kk_i^2-\frac{1}{K^2}\sum_{i=1}^K\sum_{j=1}^Kk_ik_j\nonumber\\
a_2&:=&\frac{\sum_{i=K+1}^Nk_i^2}{N-K+1}+\frac{\sum_{j=1}^Kk_j^2}{K}-2\frac{\sum_{i=1}^K\sum_{j=K+1}^Nk_ik_j}{K(N-K+1)}\nonumber\\
a_3&:=&\frac{\sum_{i=K+1}^Nk_i^2}{N-K+1}-\frac{\sum_{i=K+1}^N\sum_{j=K+1}^Nk_ik_j}{(N-K+1)^2}\nonumber
\er
and expressions $I_j(n)$ stands for some integrals. One gets
$$
I_1(n):=\int_0^{\infty}dy\frac{y^{\frac{1}{n}+1}}{(1+y)^2}=-\frac{\pi (n+1)}{n\sin\left(\frac{\pi}{n}\right)}\qquad n=-1,-2,\cdots .
$$
The integral $I_1(n)$ diverges for positive values of $n$. The integral $I_2(n)$ reads
$$
I_2(n):=\int_0^{\infty}dy\frac{y^{\frac{1}{n}}}{(1+y)^2}=\frac{\pi }{n\sin\left(\frac{\pi}{n}\right)}\qquad n=\pm 2,\pm3,\cdots .
$$
where divergence occurs only for $n=\pm1$. The last integral is of the form 
$$
I_3(n):=\int_0^{\infty}dy\frac{y^{\frac{1}{n}-1}}{(1+y)^2}=\frac{\pi(n-1) }{n\sin\left(\frac{\pi}{n}\right)}\qquad n=1,2,\cdots .
$$
One can see that there are not values of $n$ such that all integrals converge simultaneously. For instance, the energy $E_2$ is finite for $n>1$ if $a_1=0$ or for $n<-1$ when $a_3=0$. The energy $E_3$ is given by the formula
\br
E_3=\frac{32\pi}{3}M^2|n|\psi'^2\left[(\beta e^2-1)(2a_1+a_2+2a_3)+(\gamma e^2-1)a_4^2\right]\nonumber
\er
where the coefficient $a_4$ reads
$$
a_4:=\frac{1}{K}\sum_{i=1}^Kk_i-\frac{1}{N-K+1}\sum_{i=K+1}^Nk_i.
$$
The term $E_4$ takes the form
\br
E_4&=&\frac{32\pi}{3}M^2(\beta e^2+\gamma e^2-2)c^{\frac{2}{n}}\frac{n^4}{|n|}\left[\frac{N-K+1}{K}\right]^{-\frac{1}{n}}\int_0^{\infty}dy\frac{y^{1-\frac{1}{n}}}{(1+y)^4}\nonumber\\
&=&\frac{16\pi^2}{3}M^2(\beta e^2+\gamma e^2-2)c^{\frac{2}{n}}\left[\frac{N-K+1}{K}\right]^{-\frac{1}{n}}\frac{n^2-1}{\sin\left(\frac{\pi}{|n|}\right)}.\nonumber
\er
This contribution to the energy does not appear in the models without potential where $\beta e^2+\gamma e^2-2=0$. The terms $\mathcal{H}_5, \mathcal{H}_6$ 
do not contribute to a total energy since they vanish for a holomorphic solution. The last contribution which comes from the potential term reads
\br
E_7&=&2\pi M^2\tilde\mu^2\int_0^{\infty}d\rho\rho V\nonumber\\
&=&\pi M^2\tilde\mu^2\frac{Q}{|n|}\frac{\left[\frac{N-K+1}{K}\right]^{-\frac{1}{n}}}{K^2(N-K+1)^2}c^{-\frac{2}{n}}\int_0^{\infty}dy\frac{y^{1-\frac{1}{n}}}{(1+y)^4}\nonumber
\er
where the function $Q$ is taken for $|u_j|^2=1$. Considering that $\tilde\mu^2$ is given by (\ref{mu}) one gets that $E_7=E_4$. It follows that for an exact solution  the potential term  and the quartic term proportional to $\beta e^2+\gamma e^2-2$ contribute equally to the total energy.

\begin{table}[t]
\begin{center}
\begin{tabular}{ccccccccc}\hline\hline 
$(n_1,n_2)$ & $E$ & $E_1$ & $E_2$ & $E_3$ & $E_4$ & $E_5$ & $E_6$ & $E_7$ \\ \hline 
$(3,~~~1)$&~~~224~~~&~18.9~&~70.0~&~86.4~&~22.6~&~0.23~&~2.93~&~22.7 \\
$(2,-1)$ &~~~226~~~&~18.9~&~71.7~&~86.5~&~22.6~&~0.22~&~2.97~&~22.8 \\\\
$(4,~~~2)$&~~~321~~~&~25.4~&~84.5~&~127~&~34.3~&~0.60~&~15.2~&~34.5 \\
$(4,~~~1)$&~~~276~~~&~25.3~&~81.4~&~112~&~31.4~&~0.59~&~-6.46~&~31.8 \\
$(3,-1)$ &~~~277~~~&~25.3~&~82.1~&~112~&~31.3~&~0.59~&~-6.35~&~31.9 \\
$(2,-2)$ &~~~328~~~&~25.3~&~90.4~&~127~&~34.2~&~0.59~&~15.5~&~34.6 \\\\
$(5,~~~2)$ &~~~378~~~&~31.5~&~93.4~&~158~&~42.7~&~0.28~&~9.40~&~42.5 \\
$(5,~~~1)$&~~~325~~~&~31.8~&~89.8~&~126~&~41.8~&~1.53~&~-8.99~&~42.9 \\
$(4,-1)$&~~~324~~~&~31.8~&~89.2~&~126~&~41.8~&~1.51~&~-8.76~&~42.9 \\
$(3,-2)$ &~~~379~~~&~31.5~&~94.8~&~158~&~42.5~&~0.28~&~9.78~&~42.7 \\
\hline\hline
\end{tabular}
\caption{\label{tab:spectra}The energy and the components (in unit of $4M^2$) 
with the potential (\ref{potcp2++}),(\ref{potcp2+-}) of $(a,b)=(0,2)$. 
The parameters are $(\beta e^2,\gamma e^2,\mu^2,k_1,k_2)=(2.0,2.0,1.0,1.0,2.0)$.}
\end{center}
\end{table}

\section{The numerical study for the nonholomorphic vortices}
In present section we study the problem of solutions of the extended $CP^N$ Skyrme-Faddeev model 
without reduction and with presence of the potential. We shall propose the form of the potential 
such that one can compute solutions for arbitrary set of integers $(n_1,n_2,\ldots,n_N)$ which 
appear in the ansatz (\ref{ansatz}). Such solutions are non-holomorphic ones and therefore they 
can be obtained as the result of numerical integration. For the numerical study, it is more 
convenient to use a new radial coordinate $y$, 
defined by $\rho=\sqrt{\frac{1-y}{y}}$. Accordingly we adopt profile functions $g_i$, 
instead of using $f_i$. The ansatz is then
\br
u_i=\frac{1}{\sqrt{N}}\sqrt{\frac{1-g_i(y)}{g_i(y)}}
e^{i(n_i\varphi+k_i\psi(w))}
\label{solution}
\er
where $\psi(w)$ is a real function of the light-cone coordinate $w\equiv z\pm t$. 
The factor $1/\sqrt{N}$ is introduced in order that the solution naturally reduces 
that of $CP^1$ when all integers $n_i$ are equal. It is worth mentioning that in this section we are not interested in reduction itself, however, for the case of reduction one can test the numerical solution comparing it with analytical one.

The equation of motion (\ref{equationr}) can be written as
\begin{eqnarray}
&&\frac{g_i''}{g_i(1-g_i)}+\Bigl(\frac{C_1'}{C_1}+\frac{1-2y}{y(1-y)}\Bigr)\frac{g_i'}{g_i(1-g_i)}+\frac{1}{2}\frac{4g_i-3}{(g_i(1-g_i))^2}g_i'^2 \nonumber \\
&&+\frac{1}{\sqrt{y(1-y)^3}}\Bigl(\frac{C_2'}{C_1}+\frac{1}{2}\frac{C_2}{C_1}\frac{1}{y(1-y)}\Bigr)n_i+\frac{1}{2}\frac{1}{y(1-y)^3}\frac{C_3}{C_1}n_i^2 \nonumber \\
&&+\frac{\prod_{k=1}^N g_k}{\Theta}\Bigl\{\sum_{l=1}^N\Bigl(\frac{g'_l}{g_l^2}\Bigr)\frac{g_i'}{g_i(1-g_i)}-\frac{1}{y(1-y)^3}\frac{C_3}{C_1}
\sum_{l=1}^N\Bigl(n_l\frac{1-g_l}{g_l}\Bigr)n_i\Bigr\} \nonumber \\
&&-\frac{\tilde{\mu}^2}{8}\frac{1}{y^3(1-y)}\frac{1}{C_1}\frac{\Theta}{N\prod_{k=1}^N g_k}\delta V_i=0,~~~~i=1,\cdots,N
\label{eqn}
\end{eqnarray}
where $\Theta\equiv\sum_{k=1}^Ng_k$, symbols $C_k$ read
\begin{eqnarray}
&&C_1\equiv \tilde{C}_{\rho\rho} =-1+(2\gamma e^2+\beta e^2)\theta (y)+(\beta e^2-1)\frac{y}{1-y}\omega (y) \nonumber \\
&&C_2\equiv i\tilde{C}_{\rho\varphi}=-3\zeta (y) \nonumber \\
&&C_3\equiv \tilde{C}_{\varphi\varphi} ~~\nonumber \\
&&=-\frac{1-y}{y}+(\beta e^2-1)\frac{1-y}{y}\theta (y)
+(2\gamma e^2+\beta e^2)\omega (y)
\label{eqn_c}
\end{eqnarray}
and $\delta V_i$ stands for contribution from the potential i.e.
\begin{eqnarray}
\delta V_i\equiv-Ng_i^2\frac{\delta V}{\delta g_i}-\sum_{k=1}^Ng_k(1-g_k)\frac{\delta V}{\delta g_k}.
\label{dV}
\end{eqnarray}


\begin{figure*}[t]
\includegraphics[width=13cm,clip]{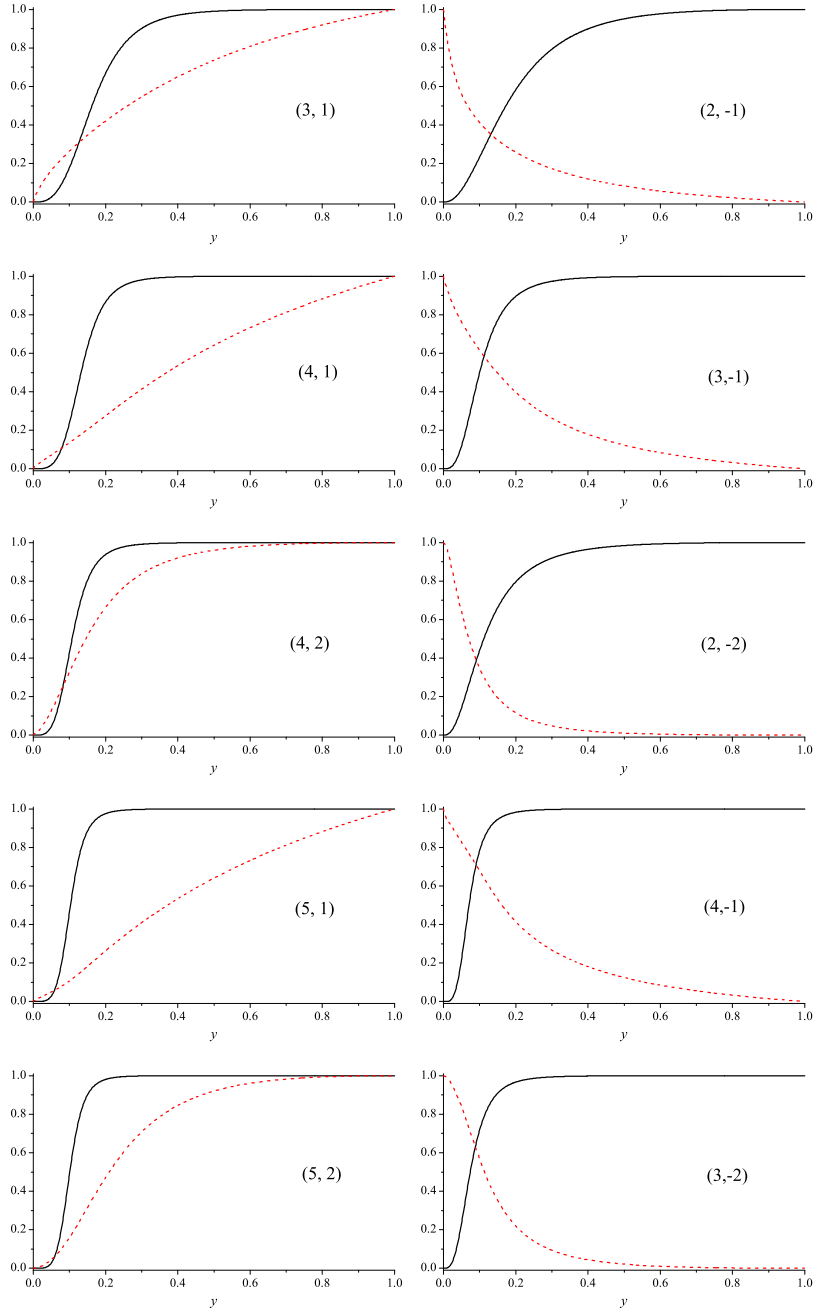}
\caption{\label{Profile}The plot of the $CP^2$ profiles $g_1$ (the solid line) and $g_2$ (the dotted line)
for the potential (\ref{potcp2++}),(\ref{potcp2+-}) of $(a,b)=(0,2)$. 
The parameters are $(\beta e^2,\gamma e^2,\tilde{\mu}^2)=(2.0,2.0,1.0)$. }
\end{figure*}

\begin{figure*}[t]
\includegraphics[width=13cm,clip]{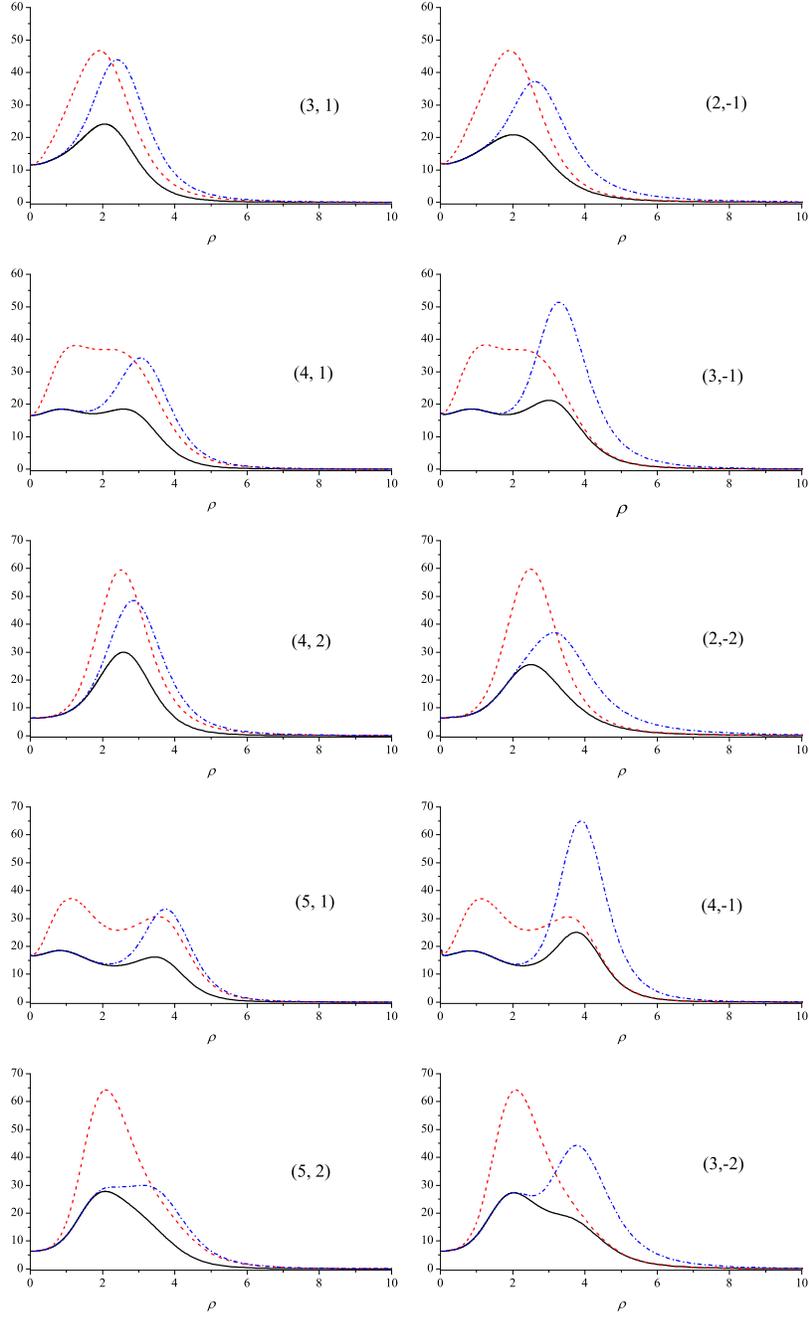}
\caption{\label{energy}The plot of the $CP^2$ density of total energies $\tilde{{\cal H}}$ (in unit of $4M^2$)
of the $(k_1,k_2)=(1.0,1.0)$ (the solid line), the $(k_1,k_2)=(1.0,2.0)$ (the dotted line) and 
$(k_1,k_2)=(2.0,1.0)$ (the dot-dashed line) for the potential (\ref{potcp2++}),(\ref{potcp2+-}) of $(a,b)=(0,2)$. 
The parameters are $(\beta e^2,\gamma e^2,\tilde{\mu}^2)=(2.0,2.0,1.0)$. }
\end{figure*}

\begin{figure}[h]
\includegraphics[width=9cm,clip]{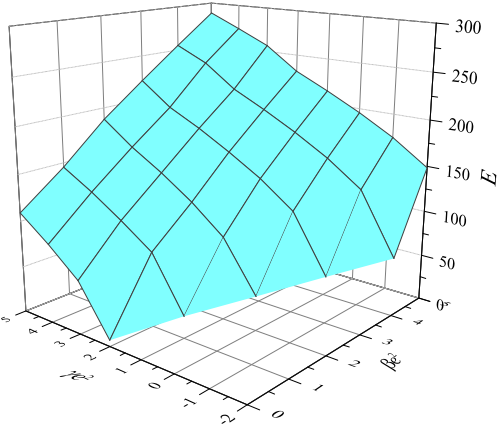}
\caption{\label{energy_bg}The total energy surface in the parameter space $(\beta e^2,\gamma e^2)$
potential (\ref{potcp2++}) of $(a,b)=(0,2)$. The remaining parameters are $(\tilde{\mu}^2,k_1,k_2)=(1.0,1.0,1.0)$. }
\end{figure}

\begin{figure*}[t]
\includegraphics[width=9cm,clip]{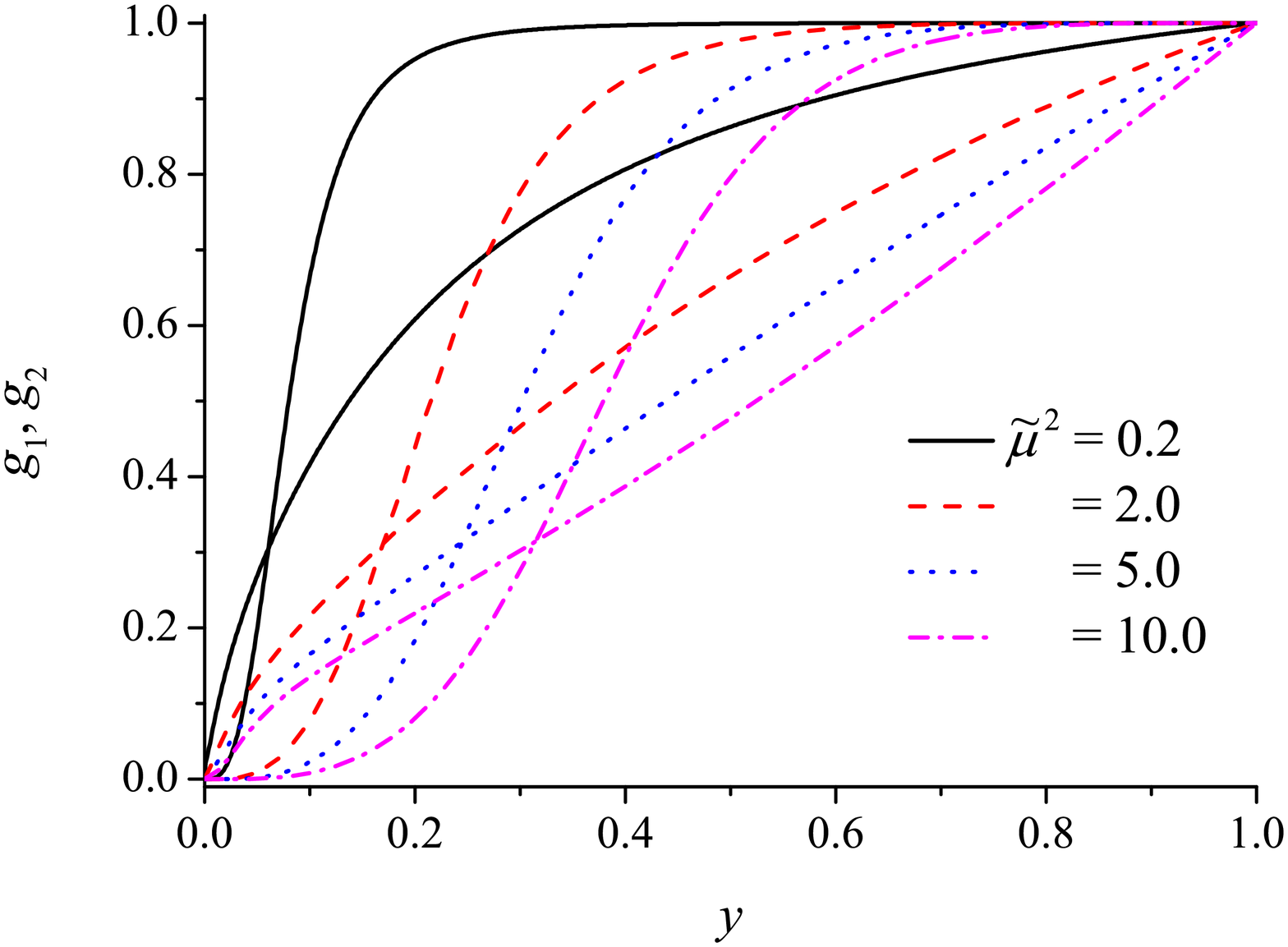}\hspace{-1.5cm}
\includegraphics[width=9cm,clip]{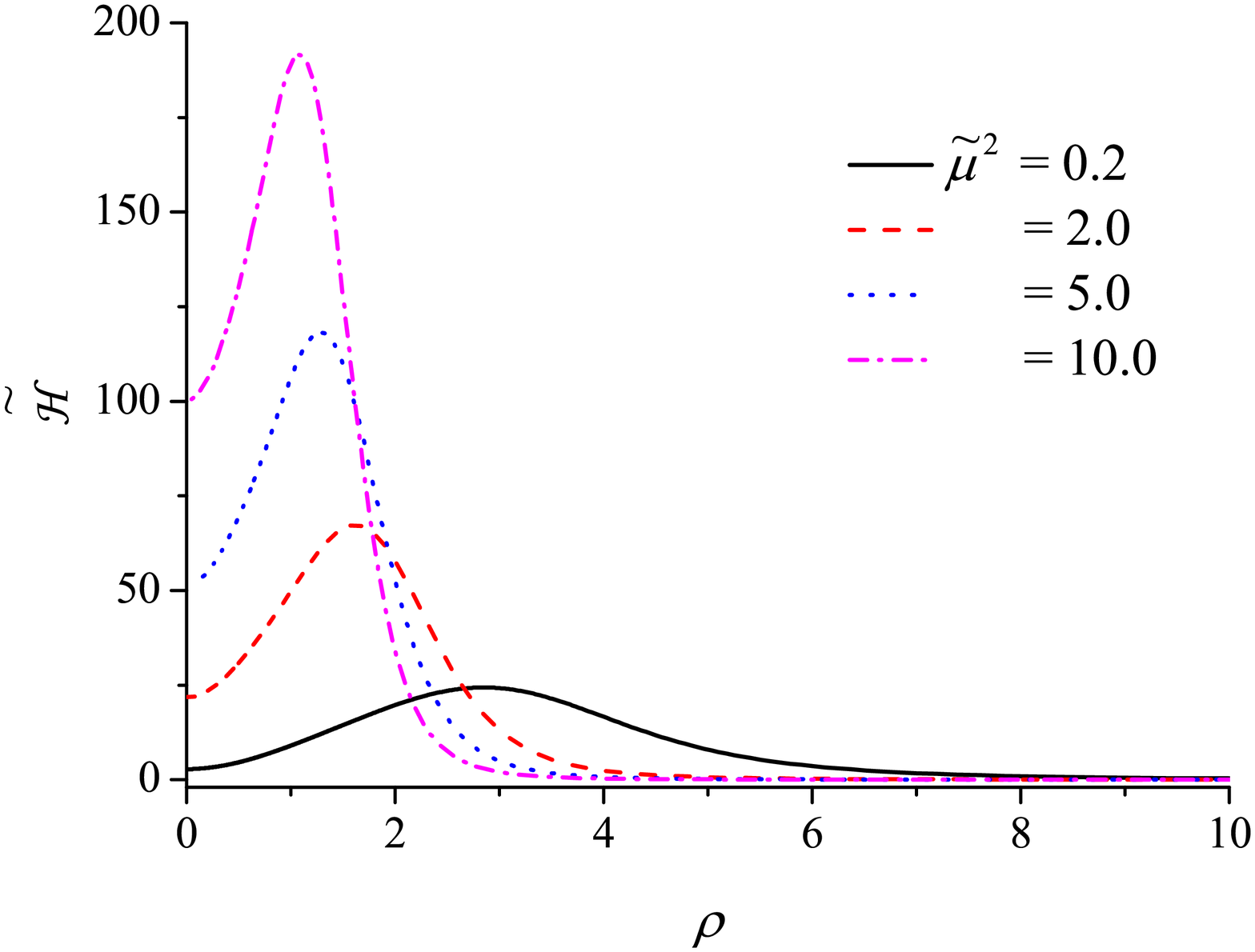}
\caption{\label{sol_mu}The profiles $(g_1,g_2)$, and the energy density $\tilde{{\cal H}}$ of the $CP^2$ 
in the potential (\ref{potcp2++}) of $(a,b)=(0,2)$ with several strength $\tilde{\mu}^2$. 
For the remaining parameters we choose $(\beta e^2,\gamma e^2,k_1,k_2)=(2.0,2.0,1.0,2.0)$. }
\end{figure*}

There is a freedom in choice of the form of the potential $V (g_i)$, however, one has to take care about 
its asymptotic behavior. A standard discussion should be based on
the vacuum structure of the field. As a simple example, we start with the $CP^1$ (O(3)) case. The O(3)
$\sigma$ model is usually defined as a vectorial triplet $\vec{n}=(n_1,n_2,n_3)$ with the constraint 
$\vec{n}\cdot\vec{n}=1$. The well-known potential named ``old-baby'' type, i.e.potential with one vacuum, 
is of the form~\cite{Piette:1994ug}
\begin{eqnarray}
V(\vec{n})=(1-\vec{n}_{\infty}\cdot\vec{n})
\label{poto3}
\end{eqnarray}
where $\vec{n}_\infty$ is a vacuum value of the field $\vec n$ at spatial infinity. 
If we choose the value $\vec{n}_\infty=(0,0,1)$ the potential becomes $V=(1-n_3)$. 
Performing stereographic projection $S^2$ on a plane we parametrize the model by a 
complex scalar field $u$ related to the triplet ${\vec n}$ by 
\begin{eqnarray}
{\vec n} = \frac{1}{1+|u|^2}\, \(u+u^*,-i\(u-u^*\),|u|^2 -1\)
\lab{udef}
\end{eqnarray}
and rewrite the potential in terms of a complex field $u$ 
\begin{eqnarray}
V(u)=\biggl(1-\frac{|u|^2-1}{1+|u|^2}\biggr)=\frac{2}{1+|u|^2}\,.
\label{poto3cp1}
\end{eqnarray}
One can expect that similar argument might work for the $CP^2$. 
In order to check this hypothesis let us consider two fields $(u_1,u_2)$ 
whose behavior at the infinity is the same as for corresponding holomorphic 
solutions characterized by two integers $(n_1,n_2)$. When $n_1,n_2>0$,  
the fields behaves as $|u_1|,|u_2|\to\infty$ for $\rho\to\infty$. 
Then one can try a generalization $|u|^2\rightarrow|u_1|^2+|u_2|^2$ resulting in the potential
\begin{eqnarray}
V(u_i)=\biggl(1-\frac{|u_1|^2+|u_2|^2-1}{1+|u_1|^2+|u_2|^2}\biggr)
=\frac{2}{1+|u_1|^2+|u_2|^2}.\label{test pot}
\end{eqnarray}
There is a serious problem which such a generalization since the model with the 
potential (\ref{test pot}) has numerical solutions only for equal values of integers $n_1=n_2$.  

A better approach to the problem is based on the observation that the potential 
in the $CP^1$ case can be expressed in terms of the $SU(2)$ valued field $U:=\vec{\tau}\cdot\vec{n}$ which allows to write (\ref{potcp1a}) as
\begin{eqnarray}
V(U)=\frac{1}{2}{\rm Tr}(1-U_{\infty}^{\dagger}U).
\label{potcp1}
\end{eqnarray}
This formula can be  easily verified using the identity 
$(\vec{\tau}\cdot\vec{n}_\infty)(\vec{\tau}\cdot\vec{n})=\vec{n}_\infty\cdot\vec{n}+i\vec{\tau}\cdot (\vec{n}_\infty\times\vec{n})$.
We examine the construction of the potential for the $N>1$ of the $CP^N$ case in the same way. A parametrization of the model, which includes the case of $SU(2)/U(1)=CP^1$ target space, is performed in variable $X$ instead of the $SU(2)$-valued field $U$. The principal variable is a function of one complex field $u$ and it reads
\begin{eqnarray}
X^{CP^1}=\frac{1}{1+|u|^2}
\left(
\begin{array}{cc}
1-|u|^2 & 2iu \\
2iu^* & 1-|u|^2  \\
\end{array}
\right)
\end{eqnarray}
or in terms of components of the unit iso-vector $\vec n$ 
$$
X^{CP^1}=-n^3 I+in^1\tau_1-in^2\tau_2,
$$
which is clearly different form $U$. 
Inverse of the principal variable $X^{CP^1}$ goes to $({X_\infty^{CP^1}})^{-1}={\rm diag}(-1,-1)$.  It follows that the following expression 
\begin{eqnarray}
V(u)=\frac{1}{2}{\rm Tr}(1-({X^{CP^1}_\infty})^{-1} X^{CP^1})&=&\frac{1}{2}{\rm Tr}(1+X^{CP^1}) \nonumber \\
&=&\frac{2}{1+|u|^2}
\label{potcp1a}
\end{eqnarray}
reproduces the potential (\ref{poto3cp1}). 
The last result constitute an important clue how to choose potentials $V(X)$ for $N>1$.

\subsection{The $CP^2$ solution}
First we give a definition of functions which still were not addressed. 
For the $CP^2$, the functions $\theta,\omega,\zeta$ in Eq.(\ref{eqn_c}) have the form
\begin{eqnarray}
&&\theta (y)=-\frac{8}{\Theta^2}y^3(1-y)\biggl[\frac{(1+g_2)g_2g_1'^2}{2g_1(1-g_1)}
\nonumber \\
&&\hspace{3cm}+\frac{(1+g_1)g_1g_2'^2}{2g_2(1-g_2)}-g_1'g_2'\biggr] \nonumber 
\end{eqnarray}
\begin{eqnarray}
&&\omega (y)=-\frac{8}{\Theta^2}\biggl[n_1^2g_1(1-g_1)g_2^2+n_2^2g_1^2g_2(1-g_2)
\nonumber \\
&&\hspace{1.5cm}+\frac{1}{2}(n_1-n_2)^2g_1(1-g_1)g_2(1-g_2)\biggr] \nonumber 
\end{eqnarray}
\begin{eqnarray}
&&\zeta (y)=-\frac{8}{\Theta^2}\sqrt{y^3(1-y)}\biggl[n_1g_2^2g_1'+n_2g_1^2g_2'
\nonumber \\
&&\hspace{1cm}+\frac{1}{2}(n_1-n_2)\bigl\{g_2(1-g_2)g_1'-g_1(1-g_1)g_2'\bigr\}\biggr]
\label{eqn_t}
\end{eqnarray}
where $\Theta=g_1+g_2$.

Generally speaking, a potential can be deduced from the asymptotic structure of solutions 
of the model. Moreover, the potential has to have the form such that the model 
has solutions for all qualitatively different combinations of the integers $(n_1,n_2)$. 
In the following subsections, we give an explicit form of the potentials for basic combinations 
of $(n_1,n_2)$.
The most crucial point is that we shall explore such potentials of which the solutions share the asymptotic behavior 
with the holomorphic counterpart, i.e. $\sim (\rho^{n_1},\rho^{n_2})$ for the combinatation $(n_1,n_2)$. 

\subsubsection{The case: $n_1>n_2>0$}
By assuming that the solution and its holomorphic counterpart have the same asymptotic behaviour at the spatial 
infinity one gets that inverse of the principal variable $X$ goes to ${X_{\infty}}^{-1}:={\rm diag}(-1,1,-1)$ as $\rho\to\infty$. 
It follows that generalization of the formula (\ref{potcp1a}) from $N=1$ to $N=2$, gives the following expression for the potential 
\begin{eqnarray}
V(u_i)={\rm Tr}(1-{X_{\infty}}^{-1}X)=4\frac{1+|u_2|^2}{1+|u_1|^2+|u_2|^2}.
\label{potcp02}
\end{eqnarray}
Note that for $\rho\to 0$ inverse of the principal variable goes to $X_0^{-1}:={\rm diag}(1,1,1)$, then the 
expression ${\rm Tr}(1-X_0^{-1}X)$ can be included as the ``new-baby'' potential which has 
two vacua~\cite{Kudryavtsev:1997nw}. 
Finally, the following expression can be considered as a general form of the potential
\begin{eqnarray}
V&=&[{\rm Tr}(1-X_0^{-1}X)]^a [{\rm Tr}(1-X_\infty^{-1} X)]^b \nonumber \\
&=&\frac{(|u_1|^2+|u_2|^2)^a(1+|u_2|^2)^b}{(1+|u_1|^2+|u_2|^2)^{a+b}} \nonumber \\
&=&\frac{(g_1+g_2-2g_1g_2)^ag_1^b(1+g_2)^b}{(g_1+g_2)^{a+b}}
\label{potcp2++}
\end{eqnarray}
where the integers $a,b$ satisfy $a\geqq 0,b>0$. 

\subsubsection{The case: $n_1>0>n_2$}
Assuming that for $n_2<0$ the field $u_2$ behaves at zero as its holomorphic counterpart i.e. 
$\sim\rho^{n_2}$ one gets that it tends to diverge as $\rho\to 0$. 
Then inverse of the principal variable $X$ goes to $X_0^{-1}:={\rm diag}(1,-1,-1)$ as $\rho\to 0$. 
The general form of the potential takes the form
\begin{eqnarray}
V&=&\frac{(1+|u_1|^2)^a(1+|u_2|)^b}{(1+|u_1|^2+|u_2|^2)^{a+b}} \nonumber \\
&=&\frac{g_1^bg_2^a(1+g_1)^a(1+g_2)^b}{(g_1+g_2)^{a+b}}
\label{potcp2+-}
\end{eqnarray}
where the integers satisfy $a\geqq 0,b>0$. 

\subsubsection{The case: $n_1,n_2<0,~~|n_1|>|n_2|$}

The asymptotic values of inverse of the principal variable are given by constant matrices 
$X_\infty^{-1}={\rm diag}(1,1,1)$ and  $X_0^{-1}={\rm diag}(-1,1,-1)$. Then the potential is 
\begin{eqnarray}
V&=&\frac{(1+|u_2|^2)^a(|u_1|^2+|u_2|^2)^b}{(1+|u_1|^2+|u_2|^2)^{a+b}} \nonumber \\
&=&\frac{g_1^a(1+g_2)^a(g_1+g_2-2g_1g_2)^b}{(g_1+g_2)^{a+b}}
\label{potcp2--}
\end{eqnarray}
where the integers satisfy $a\geqq 0,b>0$.

There still be a freedom for choice of the parameters $(a,b)$. Here we consider the simplest case, 
i.e., $(a,b)=(0,2)$  for which the forms (\ref{potcp2++}) and (\ref{potcp2+-}) become identical. 
(Note that from the asymptotic analysis there are no solution for $b=1$.) 
From (\ref{dV}), the contributions of the potential term $\delta V_i$ can easily be estimated as
\begin{eqnarray}
\delta V_1= -4\frac{g_1^2g_2(1+g_2)}{(g_1+g_2)^{2}},~~\delta V_2= 0.
\label{contv2}
\end{eqnarray}

\subsubsection*{Expansions}

We examine the asymptotic behavior of the solutions expanding the equations with (\ref{contv2}). 
the asymptotic behavior at the spatial infinity $\rho\rightarrow\infty$ ($y=0$) is given by the series expansion
\begin{eqnarray}
&&\hspace{1.0cm}g_1(y)=c_1y^{n_1}+O(y^{n_1+1}) \nonumber \\
&&\hspace{1.0cm}g_2(y)=c_2 y^{n_2}+O(y^{n_2+1}) 
\end{eqnarray}
for $n_1,n_2>0$ and 
\begin{eqnarray}
&&\hspace{1.0cm}g_1(y)=c_1 y^{n_1}+O(y^{n_1+1}) \nonumber \\
&&\hspace{1.0cm}g_2(y)=1+c_2y^{n_2}+O(y^{n_2+1}) 
\end{eqnarray}
for $n_1>0,n_2<0$. The $c_i$ is arbitrary constants (``shooting parameters'') and then all the higher order coefficients can be written by $c_i$. 
It has been also checked for the present potential that expanded solution has good asymptotic behavior at the origin $\rho\to 0~(y\to 1)$.

\begin{figure*}[t]
\includegraphics[width=9cm,clip]{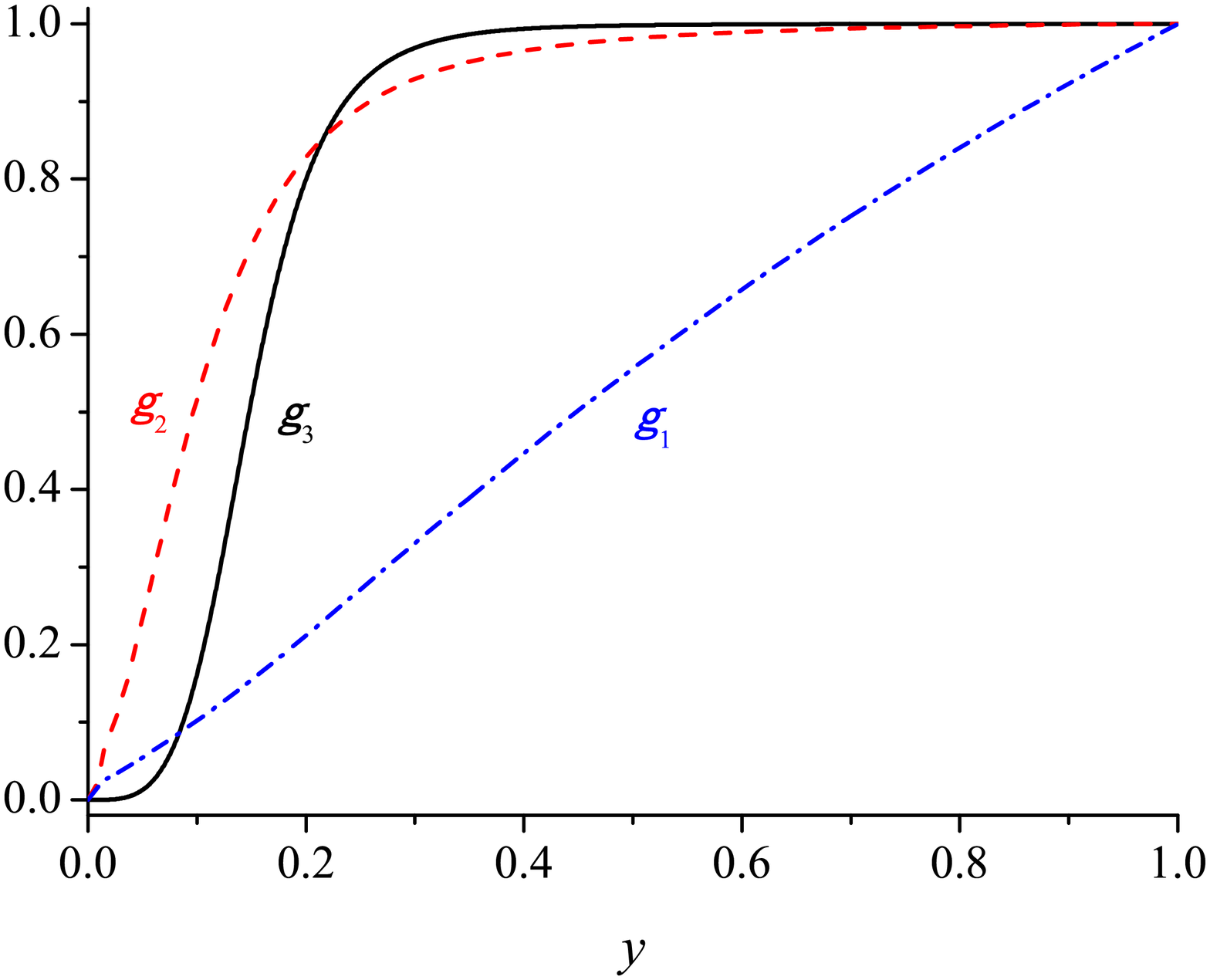}\hspace{-1.5cm}
\includegraphics[width=9cm,clip]{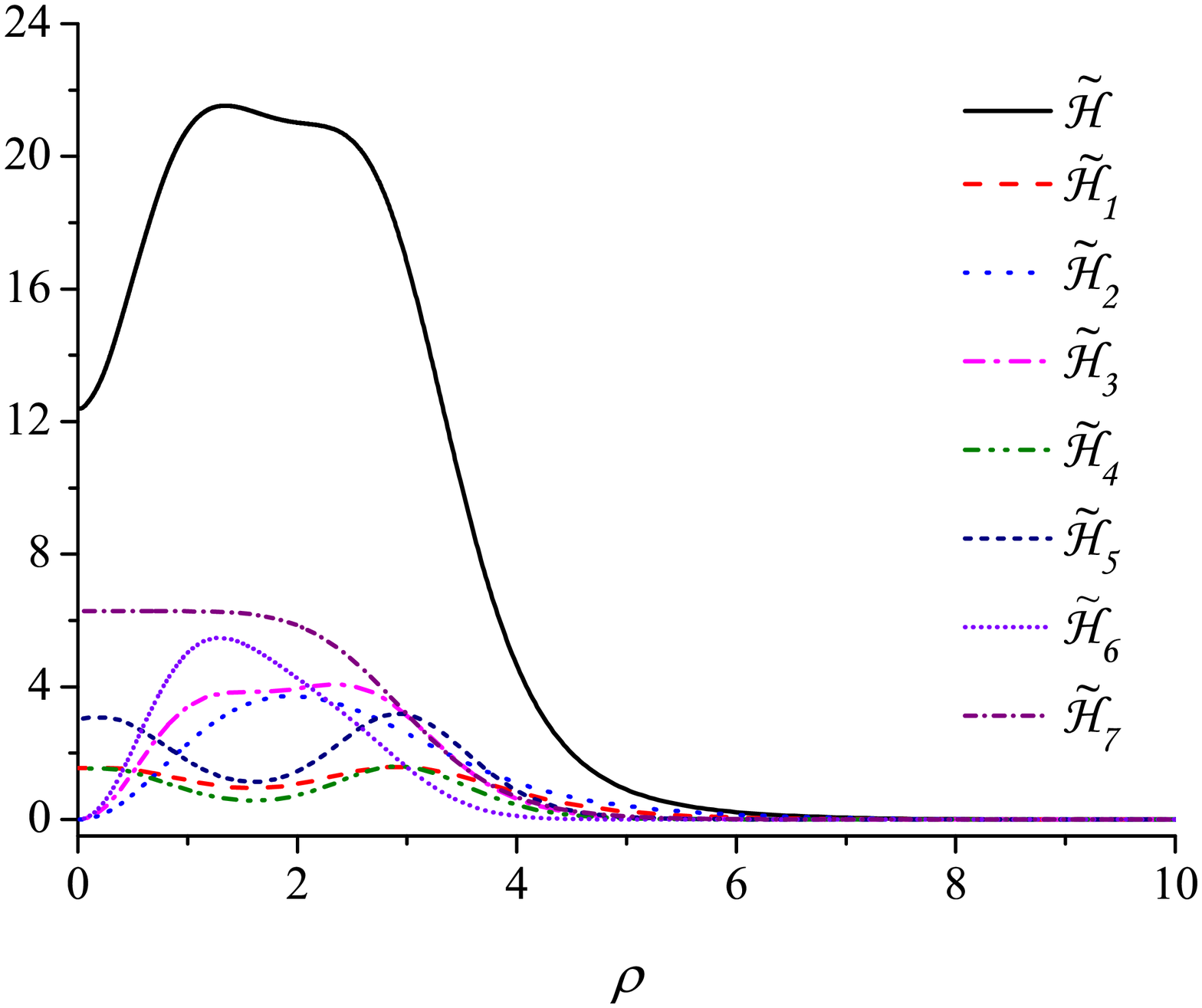}
\caption{\label{sol_cp3}The profiles $(g_1,g_2,g_3)$, and the energy, their components of the $CP^3$ 
in the potential (\ref{potcp3}) of $(a,b)=(0,2)$. 
For the remaining parameters we choose $(\beta e^2,\gamma e^2,\tilde{\mu}^2,k_1,k_2,k_3)=(2.0,2.0,1.0,1.0,2.0,1.0)$. }
\end{figure*}

The fact that our numerical solutions and holomorphic solutions have the same leading asymptotic behavior
means that they share the problem of convergence of the energy contributions. The analysis performed in 
\cite{fk} shows that not all combinations of integers $(n_1, n_2)$ leads to finite energy per unit of length.  The most 
troublesome term is ${\cal H}_2$. For this reason we shall study its asymptotic behavior. For instance, 
the term ${\cal H}_2$ has the leading expansion term around
$y\sim 0$ (for the case of $\psi(w)=w$)
\begin{eqnarray}
{\cal H}^{(2,1)}_2=\frac{8c_1 (k_1-k_2)^2}{c_2}y+O(y^2)
\end{eqnarray}
which causes the divergence of the integral unless $k_1=k_2$
because the integral
\begin{eqnarray}
\int d^2x {\cal H}^{(2,1)}_2=2\pi\int_0^1 \frac{dy}{2y^2}{\cal H}^{(2,1)}_2
\end{eqnarray}
has logarithmic divergence at $y=0$. 
For the cases such as $(n_1,n_2)=(3,1)$ the density is  
\begin{eqnarray}
{\cal H}^{(3,1)}_2=\frac{8c_1 (k_1-k_2)^2}{c_2}y^2+O(y^3).
\end{eqnarray}
One can easily see that it leads to finite energies per unit of length. 
In the following part, we will present results only for non-divergent cases. 

The numerical analysis is performed by a standard relaxation technique of which a typical mesh size is chosen as $N_{\rm mesh}=1000$, 
which supports the good convergence property of the solution. 
In Table \ref{tab:spectra}, we summarizes values of the energy and the components. The $E_1$ is the topological term, i.e.
its value is $2\pi n_1 $ (for $n_2>0$) or $2\pi (n_1+|n_2|)$ (for $n_2<0$). Our results are qualitatively good and the 
uncertainty is less than 1 percent. 
Again note that the Derrick's scaling argument for two spatial dimensions implies that the energy per unit 
length from the quartic terms and the 
potential terms should balance,i.e., $E_4+E_5=E_7$. Then, the maximal vlaue of the uncertainty  of our numerical 
results is $\sim 1$ percent.
Fig.\ref{Profile} plots the several profile functions. 
Now we define the hamiltonian densities $\tilde{{\cal H}}_k$ in terms of the energy per unit length  
\begin{eqnarray}
E_k=2\pi M^2\int_{0}^{\infty} \rho d\rho {\cal H}_k=4 M^2\int_{0}^{1} \frac{dy}{2y^2}\tilde{{\cal H}}_k.
\end{eqnarray}
In Fig.\ref{energy} we present the $\tilde{{\cal H}}$ as a function of radial coordinate $\rho$ for some values of 
$(k_1,k_2)$.    
Fig.\ref{energy_bg} shows the total energy surface in the model parameter space $\beta e^2$ and $\gamma e^2$. 

When $\beta e^2+\gamma e^2=2$, the holomorphic solutions (\ref{holomorphic}) are scale invariant, then the coefficients
are freely chosen. The fourth order terms proportional to i.e. $\beta e^2+\gamma e^2\neq 2$, breaks the scale invariance
of the model what leads to the fixing of the coefficients at some values. 
It turns out that the presence of only such fourth order terms does not lead to numerically stable solutions. 
In order to find solutions, we introduce the potential which fixes the solution corresponding to 
the highest integer $n_k$. As a consequence, the Derrick's theorem is satisfied because the coefficients of all remaining
components are properly determined.  
It is worth to examine the behavior of the size of solutions as a function of the strength of the potential $\tilde{\mu}^2$. 
In Fig.\ref{sol_mu}, we show the plot; one can see that when increasing $\tilde{\mu}^2$  the solutions became better localized around the center.   

\subsection{The higher $N$ solutions}
A generalization of the presented approach for the $N>2$ is almost straightforward. 
Now we consider the case of all integers are positive, i.e. $n_i>0,i=1,\cdots,N$. 
If $u_1$ has the highest positive integer, inverse of the principal variable $X$ goes to ${X_{\infty}}^{-1}=(-1,1,\cdots,1,-1)$, 
then the general form of the potential is then 
\begin{eqnarray}
V(u_i)=\frac{(\sum_{i=1}^N|u_i|^2)^a(1+\sum_{j=2}^N|u_j|^2)^b}{(1+\sum_{k=1}^N|u_k|^2)^{a+b}}
\end{eqnarray}
with $a\geqq 0,b>0$. 
If the two highset positive integers are equal $n_1=n_2$, the potential naturally reduces to the $CP^{N-1}$. 
We examine the case of the $CP^3$. Inverse of the principal variable $X$ goes to 
\begin{eqnarray}
X_\infty^{-1}=
\left(
\begin{array}{cccc}
0 & -1 & 0 & 0 \\
-1& 0  & 0 & 0 \\
0 & 0 & 1 & 0 \\
0 & 0  & 0 & -1 \\
\end{array}
\right)
\end{eqnarray}
which results in form of the potential
\begin{eqnarray}
V(u_i)&=&{\rm Tr}(1-X_{\infty}^{-1}X) \nonumber \\
&=&2\frac{2+|u_1-u_2|^2+2|u_3|^2}{1+|u_1|^2+|u_2|^2+|u_3|^2}.
\end{eqnarray}
When we put $u_1=u_2=u/\sqrt{2}$, the result coincides exactly with the result obtained for $CP^2$ (\ref{potcp02}).
Here we examine the case of which $n_1>n_2>n_3>0$ and 
the expressions $\delta V_i$ have the following form 
\begin{eqnarray}
&&\delta V_1= -6\frac{g_1^2g_2g_3(g_2+g_3+g_2g_3)}{(g_2g_3+g_3g_1+g_1g_2)^{2}}\nonumber \\
&&\delta V_2= \delta V_3=0.
\label{potcp3}
\end{eqnarray}
A typical result of the $CP^3$ is shown in Fig.\ref{sol_cp3}. The value of the 
topological term is $E_1\sim 25.3$ and the combination of energies per unit length 
concerned with the Derrick's becomes $E_4+E_5-E_7\sim 0.12$.

\section{Summary}

In the present paper we were focusing on the problem of solutions of the extended $CP^N$ Skyrme-Faddeev model
for wide range of coupling constants i.e. when $\beta e^2+\gamma e^2\neq 2$. The results of our analytical and numerical studies
indicates that such solutions do exist, however, they require a presence of the potential term. 
In the first part we have considered some special cases when the reduction from the $CP^N$ to the $CP^1$ happens. 
A particular choice of the potentials in such cases enabled us to obtain the   
holomorphic solutions in the sector of coupling constants $\beta e^2+\gamma e^2\neq 2$. 
There is still an open problem if such solutions do exist for non-reduced $CP^N$ case.

For a general case a freedom of choice of the potential is much larger. As a consequence
our potentials consist of one vacuum type (``old-baby'' potential) as well as of two vacua type (``new-baby'' potential). 
Both potentials break the scale invariance of the solutions and then they satisfy the Derrick's theorem. 
The numerical solutions presented in the paper does not satisfy the zero curvature 
condition and therefore they are not holomorphic functions, however, they have 
the same asymptotic behavior as its holomorphic counterparts characterized by $n_k$.

In this paper, we have numerically examined only the solutions with the rotational symmetry in the case of the old-baby type 
potential. The new-baby potential should have the similar solutions, too. 
It is a well-known fact that the old-baby potential tends to split solutions with topological charge $B$ into $B$ independent fractions, while the new-baby type does not~\cite{Hen:2007in}. It would be interesting to investigate if such behavior do really manifest for non-central vortex solutions in the model with our potentials. Such a study is important since it can serve as a test for validity of our proposal for that potential. The analysis of this subject is in progress and the results will be reported in near future.

{\bf Acknowledement}

The authors would like to thank L. A. Ferreira for discussions and comments.
We are grateful to Kouichi Toda for useful discussions. 
This work was financially supported by a grant of Heiwa Nakajima Foundation 
especially for stay of Pawe\l~Klimas in Japan.

\end{document}